\documentclass[final]{aipproc}

\layoutstyle{6x9}

\newcommand{\povo}{Dipartimento di Fisica, Universit\`{a} di Trento,
and I.N.F.N., Gruppo di Trento, Povo (TN), Italy}
\newcommand{\CQG}{Class. Quant. Grav.}
\newcommand{\PRL}{Phys. Rev. Lett.}

\newcommand{\PRD}{Phys. Rev. \bf{D} }
\newcommand{\etal}{\textit{et al.}}
\newcommand{\fref}[1]{fig.\ref{#1}}

\begin{document}

\title{Torsion pendulum facility for direct force measurements of LISA
GRS related disturbances}


\author{L. Carbone}{address=\povo, email={carbone@science.unitn.it}}
\author{A. Cavalleri}{address={Centro Fisica degli Stati Aggregati, Povo (TN),
Italy}}
\author{G. Ciani}
{address=\povo}
\author{R. Dolesi}
{address=\povo}
\author{M. Hueller}
{address=\povo}
\author{D. Tombolato}
{address=\povo}
\author{S. Vitale}
{address=\povo}
\author{W. J. Weber}
{address=\povo}

\begin{abstract}
A four mass torsion pendulum facility for testing of the LISA GRS is
under development in Trento. With a LISA-like test mass suspended
off-axis with respect to the pendulum fiber, the facility allows for
a direct measurement of surface force disturbances arising in the
GRS. We present here results with a prototype pendulum integrated
with very large-gap sensors, which allows an estimate of the
intrinsic pendulum noise floor in the absence of sensor related
force noise. The apparatus has shown a torque noise near to its
mechanical thermal noise limit, and would allow to place upper
limits on GRS related disturbances with a best sensitivity of 300
fN/$\sqrt{\textrm{Hz}}$ at 1mHz, a factor 50 from the LISA goal.
Also, we discuss the characterization of the gravity gradient noise,
one environmental noise source that could limit the apparatus
performances, and report on the status of development of the
facility.
\end{abstract}

\maketitle

{\small Submitted to {\it Proc. of the $6^{ th }$ Int. LISA
Symposium}, AIP Conference Proceedings, 2006}

\section{Introduction}
The LISA gravity wave sensitivity goal requires that the residual
acceleration of its test masses (TMs) must be kept below 3
fm/s$^{2}$/$\sqrt{\textrm{Hz}}$ down to 0.1mHz \cite{bender:LISA}.
This is obtained with a Drag-Free Control, with the spacecraft that
shields the TMs from environmental disturbances and is centered
about the TMs geodesic motion according to a position sensor, also
called ``Gravity Reference Sensor'' (GRS). The current design of the
LISA GRS foresees cubic 2~kg and (46~mm)$^3$ TMs, surrounded by an
array of electrodes included in a capacitive readout and actuation
scheme \cite{weber:sensor,dolesi:sensor}. In preparation for the
flight-test of the LISA core elements, LTP \cite{vitale:LTP},
relevant upper limits on the most important GRS related surface
disturbances have been already set by means of torsion pendulums
\cite{carbone:prl,carbone:char}. These experiments used a single
LISA-like TM, hanging in axis from a torsion fiber and surrounded by
one GRS prototype: this configuration is extremely sensitive to
torques arising in the GRS but is insensitive to net forces and to
any source acting along axes passing through the center of the TM.
Upper limits on force noise sources have been thus set using
model-dependent arm-lengths, to convert the torques into forces
\cite{hueller:upperlimits}.

To overcome this limit, we are developing a new torsion pendulum
facility, employing a four TM configuration \cite{hoyle:4mass}, that
allows for a direct measurement of force disturbances arising in the
GRS. We present here results with a first prototype pendulum that
allows an estimate of the intrinsic pendulum performance. We discuss
the characterization of the gravity gradient noise, one
environmental noise source that could limit the apparatus
performances, and finally report on the status of development of the
facility.

\section{Development of the facility}
\begin{figure}[t]
\centering
  \includegraphics[width=100mm]{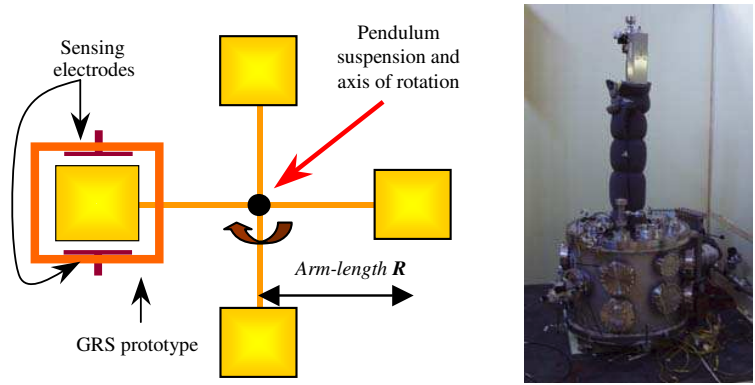}
  \caption{\label{facility:fig} On the left, a sketch of the experiment concept (top view).
  On the right, a photograph of the facility: its main component, the vacuum chamber, is visible.}
\end{figure}
A sketch of the experiment concept is shown in \fref{facility:fig},
with a photograph of the facility. Four LISA-like TMs lay on the
horizontal plane, are mounted on a cross-shaped support, and are
suspended off-axis with respect to the pendulum torsion fiber, at
distance $R$ bigger than the TM side-length. One of the TMs is
included in a GRS prototype: any net force arising inside the GRS
gets thus converted into a torque on the pendulum by the arm-length
$R$. This pendulum has full sensitivity to disturbances arising
inside the sensor, independently from their location, and the unique
conversion arm-length $R$ allows to set model-independent force
noise upper-limits. Use of hollow TMs reduces the pendulum weight
and allows for thinner torsion fibers. This makes the pendulum
mainly sensitive to surface forces (however the most relevant
envisioned by noise analyses) but increases its force sensitivity,
that can be further enhanced by enlarging the pendulum arms, in an
optimal compromise with low weight and coupling to gravity gradients
(see below). The direct measurement of GRS relevant parameters like
force gradients (``stiffness'') or stray electrostatic forces will
provide then fully representative estimates for LISA.

The main component of the facility is a 250~$l$ vacuum vessel (see
\fref{facility:fig}), accomodating the pendulum and the GRS, that
sits on a concrete slab partially isolated from the laboratory
floor. A 90~cm long tube encloses the torsion fiber. A 400~$l/s$
turbo pump sets the residual pressure at $\approx10^{-7}$mBar. The
entire apparatus is enclosed in a thermally insulated box whose
temperature is controlled by a thermalized water bath stabilizing
the air circulating inside a heat exchanger. The experiment is
equipped with manual micro-manipulators (4 dof for GRS alignment + 2
dof for pendulum rotational/vertical alignment) and motorized stages
(2 translators for GRS alignment and automatized stiffness
measurements). The pendulum motion is monitored by a commercial
autocollimator, with 50~nrad resolution for both twist and tilt
modes. The pendulum tilt motion is suppressed with a magnetic eddy
current damper (decay time $\tau\approx 200$~s). Finally, the
facility is equipped with different environmental monitors, like
thermometers for in-vacuum and in-air temperatures, two 3-axis
flux-gate magnetometers for magnetic fields and field gradients and
a 2-axis tilt-meter for the residual tilt motion of the apparatus.

\begin{figure}[t]
\centering
  \includegraphics[height=55mm]{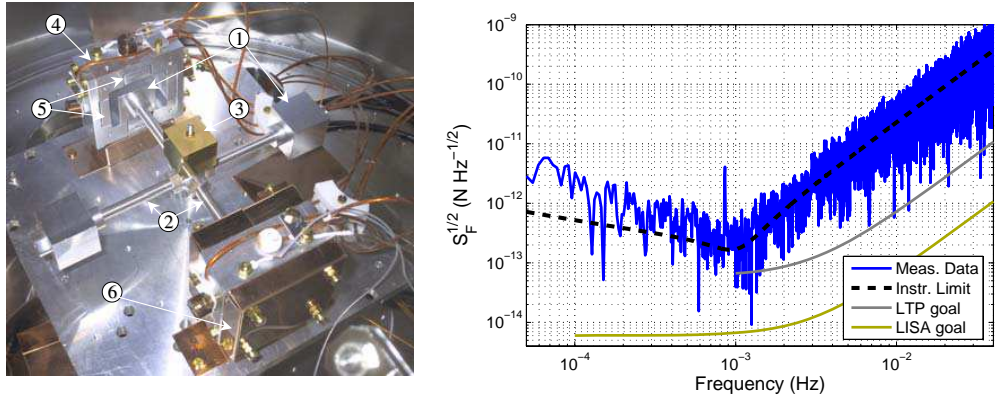}
  \caption{\label{noise:fig} Left: a photograph of the first pendulum
  prototype, in the vacuum vessel: 1. TMs 2. Cross-shaped support
  3. Mirror 4. Capacitive sensors 5. Sensing electrodes 6.
  Copper electrostatic shields.
  Right: typical force noise sensitivity of the apparatus, compared with LISA and LTP
  residual force noise goals.}
\end{figure}
As the main GRS-related noise sources are expected to depend
strongly on the TM-electrode separation (gap), an estimate of the
intrinsic pendulum noise floor has been performed with a prototype
pendulum integrated with very large-gap sensors (allowing us also to
debug the facility and verify the fiber performances). We employed
(3~cm)$^3$ solid uncoated Al TMs ($\approx 67$~g), mounted on an Al
cross shaped support and centered at a distance of $R=$89~mm from
the fiber axis (see \fref{facility:fig}). A Au-coated mirror was
mounted along the fiber axis. The total pendulum weight was $\approx
460$~g, roughly the same of the next pendulum, with (46~mm)$^3$ TMs.
Two single axis capacitive sensors surrounded two of the TMs
(electrically isolated) at very large gaps (8~mm along the force
sensitive axis), providing both a high sensitivity (200~nrad
/$\sqrt{\textrm{Hz}}$) readout of the pendulum twist motion and
actuation authority. We used a 95~cm long, 50~$\mu$m diameter W
fiber, with a measured mechanical quality factor $Q \approx 1000$.
With the pendulum inertia $I\approx 2.6 $~g m$^2$ and a torsion
constant $\Gamma\approx 80$~nNm/rad, the torsion pendulum resonance
was $\approx0.8$~mHz.

A typical force sensitivity performance of the apparatus, obtained
by dividing the measured pendulum torque noise by the arm-length
$R=89$~mm, is shown in \fref{noise:fig}. Below 1~mHz, the observed
noise is within a factor 2 from the instrumental limit, set here by
the pendulum mechanical thermal noise. Above 1~mHz, the sensitivity
is limited by the angular read-out noise. The performance shown in
\fref{noise:fig} would allow to place upper limit on the
disturbances related to a GRS prototype at the level of
800~fN/$\sqrt{\textrm{Hz}}$ between 0.2 and few mHz, with a best
sensitivity of 300~fN/$\sqrt{\textrm{Hz}}$ at 1~mHz, a factor 50
from the LISA goal. The excess noise at low frequencies, that could
be given by environmental temperature fluctuations as well as tilt
motion of the apparatus, is still being investigated.

A possible drawback of this four mass pendulum configuration is its
sensitivity to gravity gradients \cite{hoyle:4mass,Y:Su}. The
gravity gradient torque induced on a pendulum by objects at a
distance $r$ goes like $(R^l/r^{(l+1)})$, with $R$ and $l$,
respectively, the characteristic radius and degree of symmetry of
the pendulum mass distribution about the torsion axis. This pendulum
symmetry is nominally high, with $l\geq 4$. However, lower order
terms, particularly quadrupole, could enter through machining and
assembling imperfections, producing mass distribution asymmetries
and tilting of the pendulum: this enhances the pendulum sensitivity
to gravity gradients and avoids reaching its ultimate performances.

\begin{figure}[t]
\centering
  \includegraphics[height=55mm]{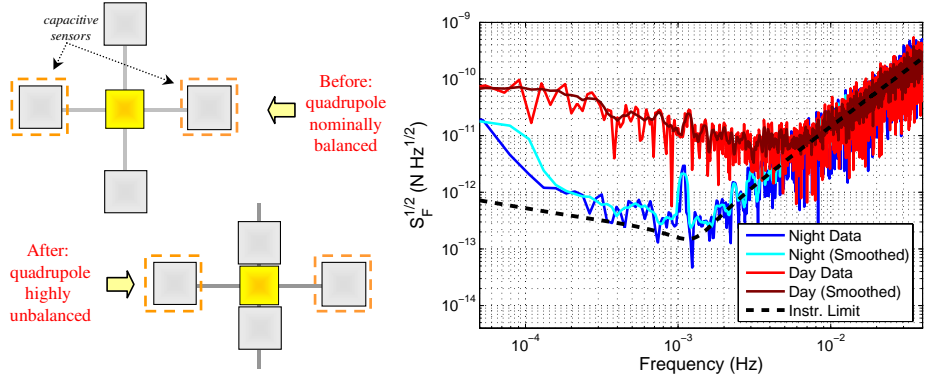}
  \caption{\label{noise:fig:nightday} Left: scheme of the enhancement of the pendulum mass quadrupole.
  Right: comparison of the
  typical night and day force sensitivities of the unbalanced quadrupole pendulum configuration.}
\end{figure}

In order to qualitatively characterize the role of gravity gradient
noise of our experimental site, we performed a test in which we
purposely enhanced the pendulum quadrupole moment ($l=2$) to detect
any induced excess noise on the pendulum. Two of the four masses
where moved towards the center of the pendulum by $\approx 50$~mm,
as shown in \fref{noise:fig:nightday}. If the previous configuration
had a ``nominally'' nulled quadrupole moment, i.e.
$Q_{x,y}\approx0$, this produced a (calculated) mass quadrupole
moment of the pendulum of order $Q_{x,y}\approx 0.96$~g~m$^2$ (for
comparison, a 1~cm displacement would produce $Q_{x,y}\approx
0.25$~g~m$^2$, and a 2~mm displacement $Q_{x,y}\approx
0.06$~g~m$^2$).

Measured force noise spectra are shown in \fref{noise:fig:nightday}.
When human activity is present close to the experiment, in
particular due to people working at few m from the apparatus, the
observed force noise is about a factor 50 higher than the
instrumental limit. However, when human activity inside and near the
lab is reduced (night and week-ends) we did not observe any excess
noise compared to the typical performance of the pendulum, as in
\fref{noise:fig}. With reasonable tolerances of the final
(46~mm)$^3$ TM assembly of order 100~$\mu$m (or better), the mass
quadrupole moment of the pendulum will be reduced at least by a
factor 300, compared to the one in \fref{noise:fig:nightday}. This
would make gravity gradient noise to be substantially negligible for
the pendulum performance. Even if this test provides only a partial
picture of the gravity gradient noise of our experimental site (this
is only one of the possible pendulum assembling imperfections, as
for instance tilting with respect to the vertical axis, for which we
did not perform a specific investigation), this test shows that
quadrupole gravity gradient noise should not represent a relevant
limitation for reaching the ultimate pendulum performance in terms
of force sensitivity.

\section{Future work: Tests with the LTP GRS}
\begin{figure}[t]
\centering
  \includegraphics[height=55mm]{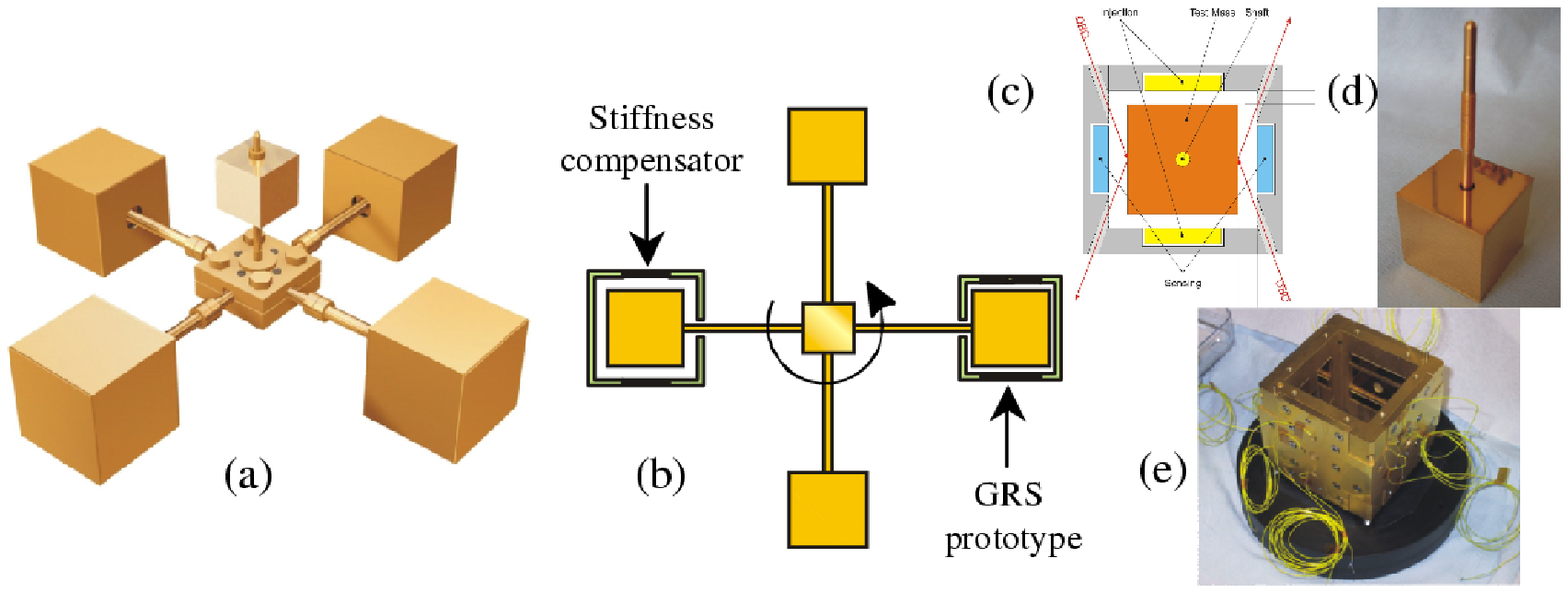}
  \caption{\label{newpend:fig} Schematic of the new pendulum,
  currently under construction: (a) a 3D cartoon of the final
  assembly; (b) top view schematic; (c) a sketch of the ``stiffness compensator'';
  (d) one of the already delivered 46mm TMs (the supporting shaft is visible); (e) a
replica of the LTP GRS Engineering Model.}
\end{figure}
We are now working on the integration of a fully representative GRS
prototype to be tested in the facility. A new pendulum, shown in
\fref{newpend:fig}, is currently under construction, employing four
Al ($46$mm)$^3$ hollow TMs. One TM will be included into a GRS
prototype that implements the LTP design. An additional capacitive
sensor, implemented with very large gaps and needed to compensate
the tilt-twist coupling induced on the pendulum by the GRS-related
stiffness, will surround the opposite TM. This ``stiffness
compensator'' will also be the support for an optical read-out
proposed for LISA by our INFN collaborators \cite{DiFiore}. These
two TMs will be electrically isolated, for the capacitive readout
and actuation. For electrostatic homogeneity, the remaining two TMs
will be electrically grounded to the rest of the pendulum and then
to the fiber. Similarly, the whole pendulum will be Au-coated,
included TMs and cross shaped support.

This facility will allow to test GRS prototypes for LTP and LISA
with full-size TMs and highly representative conditions of the
flight mission. By direct measurement of forces instead of torques,
it will allow stringent upper limits on LISA force noise, which will
be independent of the model of the noise source and its location
within the GRS. Furthermore, the role of the sources that produce
force noise on the LISA TMs will be more straightforward to evaluate
and model, leading to a more accurate verification of the LISA noise
model.

The authors would like to thank E.Adelberger and C.D.Hoyle for many
useful discussions on the gravity gradient issue. This work was
supported by ASI, ESA and INFN.

\bibliographystyle{aipproc}

\begin{thebibliography}{99}
\bibitem{bender:LISA} Bender P \etal, LISA ESA-SCI(2000)11, 2000
\bibitem{weber:sensor} Weber W J \etal 2002 {\it SPIE Proc.} {\bf 4856} 31
\bibitem{dolesi:sensor} Dolesi R \etal 2003 {\CQG {\bf 20} S99}
\bibitem{vitale:LTP} Anza S \etal 2005 {\CQG {\bf 22} S125}
\bibitem{carbone:prl} Carbone L \etal 2003 {\PRL {\bf 91}} 151101
\bibitem{carbone:char} Carbone L \etal 2005 {\CQG {\bf 22} S509}
\bibitem{hueller:upperlimits} Carbone L \etal 2004 {\CQG {\bf 21} S611}
\bibitem{hoyle:4mass} Carbone L \etal Proc. of 10$^{th}$ Marcel Grossmann Meeting on General Relativity, gr-qc/0411049
\bibitem{Y:Su} Y. Su \etal 1994 {\PRD {\bf 50}}, 3614
\bibitem{DiFiore} Di Fiore L. \etal, this conference proceedings
\end{thebibliography}

\end{document}